\documentclass[showpacs,twocolumn,floatfix,prl]{revtex4}
\usepackage{graphicx} 	
\usepackage{bm} 		
\usepackage{amssymb}
\usepackage{amsmath}
\usepackage{amsfonts}
\usepackage{epstopdf}
\usepackage{times}

\renewcommand{\section}[1]{{\par\it #1.---}\ignorespaces}
\newcommand{\swave}{$s$-wave}
\newcommand{\pwave}{$p$-wave}
\newcommand{\dwave}{$d$-wave}
\newcommand{\tildet}{$\tilde{t}$}
\newcommand{\muI}{$\mu_{\rm I}$}
\newcommand{\dxywave}{$d_{xy}$-wave}

\begin{document}

\title{Self-consistent superconducting proximity effect at the quantum spin Hall egde}

\author{Annica M. Black-Schaffer}
 \affiliation{NORDITA, Roslagstullsbacken 23, SE-106 91 Stockholm, Sweden}

\date{\today}
\begin{abstract}
We study self-consistently a microscopic interface between a quantum spin Hall insulator (QSHI) and a superconductor (SC), focusing on properties related to Majorana fermion creation. For an \swave\ SC we show that odd-in-momentum, or \pwave, order parameters exist for all doping levels of the QSHI and that they can be related to different spinless Cooper pair amplitudes. Despite this, the induced superconducting gap in the QSHI always retains its \swave\ character, validating the commonly used effective model for superconductivity in a topological insulator. For a \dxywave\ SC, we show that a Majorana mode is only created at finite doping and that there is no excitation gap protecting this mode.
\end{abstract}

\maketitle
Topological insulators (TIs) are a new class of materials where the bulk is gapped but there exist conducting surface states which are robust against all time-reversal invariant perturbations. With the experimental discovery of a two dimensional (2D) TI, or equivalently, a quantum spin Hall insulator (QSHI) \cite{Bernevig06all}, in a HgTe/CdTe heterostructure \cite{Konig07} and the subsequent advent of 3D TIs \cite{Hsieh08all}, these materials have lately gained a large amount of attention \cite{Hasan10, Qi10}. 
One of the areas attracting considerable interest is when the TI surface state is superconducting. Fu and Kane \cite{Fu08} showed that the low-energy spectrum in a 3D TI with an \swave\ superconducting order parameter at high doping levels resembles that of a spinless $p_x+ip_y$ superconductor (SC) and thus supports Majorana fermions at vortices \cite{Read00} or at the interface between superconducting and ferromagnetic regions \cite{Fu08, Fu09}. The non-Abelian statistics of a Majorana fermion can provide fault tolerant topological quantum computation \cite{Nayak08} and multiple proposals already exist on how to detect Majorana fermions in TIs \cite{MajoranaTIsall}.
However, despite the far reaching consequences of the \pwave\ character of the superconducting state, very little is known about its microscopic origins. 
For example, the Majorana mode survives even when the doping level goes to zero \cite{Fu08, Qi10} although there has so far not been any reported signs of $p_x+ip_y$-character in this doping regime.
Moreover, superconductivity is in general only added on a phenomenological level as a constant order parameter to the effective Hamiltonian of the surface TI state \cite{Fu08, Fu09, MajoranaTIsall} 
despite being generated by proximity to a SC.
Very recently Stanescu {\it et al.} \cite{Stanescu10} studied proximity induced superconductivity in a 3D TI and found, using several approximations, a \pwave\ order parameter only at high doping levels. They also predicted that this component will eventually close the superconducting gap, which would, in fact, render the commonly used phenomenological model with a constant \swave\ order parameter invalid in the high doping regime.
A comprehensive picture of the symmetries of the induced superconducting state in a TI is thus clearly lacking and, as a consequence, the connection to the existence of Majorana fermions is not well established.

In this Rapid Communication we will treat a QSHI-SC interface fully self-consistently, in which we do not only calculate the superconducting proximity effect, or the leakage of Cooper pairs into the QSHI, as has been done before \cite{Stanescu10}, but also the accompanied loss of Cooper pairs, or inverse proximity effect, on the SC side of the interface. Even more importantly, we will explicitly calculate all significant \pwave\ amplitudes and show that there are, in fact, large contributions even at zero doping. Our results demonstrate that it is natural that the Majorana mode survives even at zero doping.
Moreover, we show that the superconducting gap never closes but always retains its \swave\ character which provides a formal motivation for the commonly used phenomenological model \cite{Fu08, Fu09, MajoranaTIsall}. 
We have also self-consistently studied the interface between a QSHI and a \dxywave\ SC, since a \dxywave\ state in a TI was recently shown to support Majorana fermions \cite{Linder10}. We show, however, that in our microscopic model the Majorana mode only appears at finite doping and what there is no excitation gap protecting this mode, thus rendering it useless for quantum computation \cite{Nayak08, Sau10}.

\begin{figure}[h!]
\includegraphics[scale = 0.29]{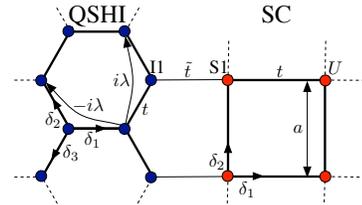}
\caption{\label{fig:setup} (Color online) Microscopic details of the QSHI-SC interface for the (1,0) surface of the SC and with sites $a_{i\alpha}$ (dark), $b_{i\alpha}$ (light).  
}
\end{figure}

To self-consistently study the interface between a SC and the edge of a QSHI we consider a minimal microscopic model defined as
$H  =   H_{\rm I} + H_{\rm S} + H_{\tilde{t}}$  
%
and schematically displayed in Fig.~\ref{fig:setup}.
%
\noindent For $H_{\rm I}$ we use the first QSHI prototype, the so-called Kane-Mele model \cite{KaneMele05}, which is defined on a honeycomb lattice with spin-orbit coupling $\lambda$:
$H_{\rm I}  =   -t \sum_{\langle i,j\rangle,\alpha} a_{i\alpha}^\dagger a_{j\alpha} + \mu_{I} \! \sum_{i,\alpha} a_{i\alpha}^\dagger a_{i\alpha} +
i \lambda \sum_{\langle \langle i,j\rangle \rangle} \nu_{ij}a_{i\alpha}^\dagger s^z_{\alpha \beta}a_{j\beta}$. 
%
Here $a$ is the fermion operator on the bipartite honeycomb lattice, $\langle i,j \rangle$ and $\langle \langle i,j \rangle \rangle$ denote nearest neighbors and next nearest neighbors respectively, $\alpha, \beta$ are the spin indices, and $\nu_{ij} = +1$ $(-1)$ if the electron makes a left (right) turn to get to the second bond. $t$ is the nearest neighbor hopping amplitude and we set $t = 1$ for simplicity. We also fix $\lambda = 0.3$ which gives a bulk band gap of 1 and thus allows the chemical potential $\mu_{\rm I}$ to vary between 0 and 1. 
We further define the SC on a 2D square lattice at half-filling with an on-site Hubbard attraction $U$ to generate a prototype conventional \swave\ superconductor:
$H_{\rm S}  =   -t\sum_{\langle i,j\rangle,\alpha} b_{i\alpha}^\dagger b_{j,\alpha} - U\sum_{i}b_{i\uparrow}^\dagger b_{i\uparrow} b_{i\downarrow}^\dagger b_{i\downarrow}$.
%
We treat $H_{\rm S}$ within mean-field theory using the self-consistency condition
$\Delta_U(i) = -U\langle b_{i\downarrow} b_{i\uparrow}\rangle$, 
%
for the \swave\ order parameter $\Delta_U$.
We will use $U = 1.2$ which gives $\Delta_U = 0.1$, but we have also verified our results for $\Delta_U = 1$.
Finally, the tunneling Hamiltonian is described by
$H_{\tilde{t}}  =   -\tilde{t}\sum_{\langle i,j\rangle,\alpha} a_{i\alpha}^\dagger b_{j,\alpha} + {\rm H.c.}$, 
%
where $a$ is on the edge site of the QSHI, I1, and b is on the first SC site, S1. \tildet\ is technically allowed to vary from 0 to 1, although we expect an experimental interface to never reach $\tilde{t} = 1$.
We align the TI so that its edge is along the zigzag direction, connect the (1,0) surface of the SC to the QSHI, and assume the same unit cell size $a$ for both the QSHI and the SC to avoid any lattice mismatch.
We also assume a smooth interface and Fourier transform in the direction along the QSHI-SC interface. 
We can solve the above model self-consistently by first diagonalizing the Hamiltonian for a guess profile of the order parameter $\Delta_U(i)$ throughout the whole structure. $\Delta_U(i)$ can then be recalculated by using the mean-field self-consistency condition. By repeating this process until $\Delta_U(i)$ does not change between two subsequent iterations, we achieve self-consistency for the superconducting state in the whole QSHI-SC structure. The superconducting proximity effect is explicitly captured in the (on-site) Cooper pair amplitude $F_U(i) = \langle a_{i\downarrow} a_{i\uparrow}\rangle$ in the QSHI and, equivalently, with $a\rightarrow b$ in the SC, where obviously also $F_U = -\Delta_U/U$. 
%
%
\section{Self-consistent superconducting gap}
\begin{figure}[htb]
\includegraphics[scale = 0.85]{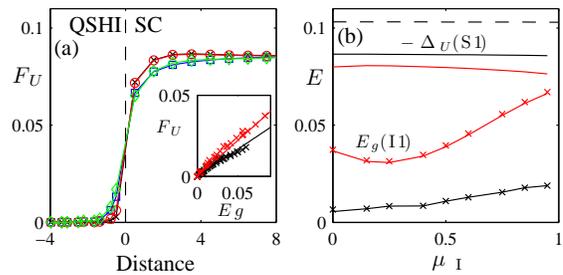}
\caption{\label{fig:PUEg} (Color online) (a): $F_U$ across the QSHI-SC interface for the combinations ($\mu_{\rm I}, \tilde{t})=  (0, 0.2)$ in (black $\times$), $(0, 0.6)$ in (blue $\square$), $(0.75, 0.2)$ in (red $\circ$), and $(0.75,0.6)$ in (green $\diamond$). Vertical dashed line marks the interface and the distance is measured in number of unit cells. Inset: $F_U$ as function of $E_g$ at site I1 for $0 \leq \mu_I < 1$, $\tilde{t}\leq 0.6$, and the bulk order parameter $\Delta_U = 0.1$ (black) and 1 (red). Crosses mark data, lines are linear fits. (b): $-\Delta_U({\rm S1})$ and $E_g({\rm I1})$ as function of \muI\ for $\tilde{t} = 0.2$ (black) and 0.6 (red). Crosses mark $E_g$(I1) reached in a non-self-consistent calculation. Dashed line indicates $|\Delta_U| = E_g$ in the SC bulk. 
}
\end{figure}

In Fig.~\ref{fig:PUEg}(a) we show the self-consistent profile of $F_U$ across the QSHI-SC interface for a few representative values of $\mu_{\rm I}$ and \tildet. As seen, there is a depletion of Cooper pairs, or inverse proximity effect, on the SC side of the junction with the accompanied leakage of pairs, or proximity effect, into the QSHI. The spread in $F_U$ on the QSHI side is essentially the same as the spread of the QSHI edge state, whereas the depletion region is a few unit cells wide on the SC side. As can be expected, the proximity effect is larger for larger \tildet, whereas, perhaps more surprisingly, there is very little dependence on the QSHI doping level \muI\ for small to moderately large tunneling amplitudes.
%
%
Only a self-consistent approach can capture the inverse proximity effect and in Fig.~\ref{fig:PUEg}(b) we plot the value of $\Delta_U$ on the first SC site, S1, as a function of \muI. As seen, this value is significantly lower than $\Delta_U = E_g$(bulk) (dashed line). 
We also in (b) plot the extracted superconducting gap in the QSHI edge state, $E_g$(I1). The self-consistent values of $E_g$ are shown with a solid line whereas non-self-consistent results, using a step-function of $\Delta_U = E_g$(bulk) at the interface, are shown with crosses and, as seen, they are essentially identical. Therefore, self-consistency is {\it not} important for calculating the induced superconducting gap at the edge of a QSHI from an \swave\ SC. This is in contrast to a \dxywave\ SC where self-consistency is crucial as we will discuss below. 
%
We also see that $E_g$(I1), in general, increases with \muI\ for all experimentally relevant \tildet. The slight downturn at small \muI\ is due to the fact that for a highly transparent interface the SC induces a slight electron doping in the QSHI edge state whereas the (positive) applied chemical potential \muI\ creates an overall hole doping. We note especially that even for the highest $\mu_{\rm I}$ we can achieve and still have a bulk gap, i.e.~$\mu_I < 1$, $E_g$(I1) is always finite. This is in contradiction to the prediction by Stanescu {\it et al.} \cite{Stanescu10} for a 3D TI. 
While \pwave\ components are induced at the interface as we show below, we attribute the non-zero $E_g$ to the fact that the \swave\ component $F_U$ on site I1 is always finite for any \muI, and thus $E_g$ is set by this value of $F_U$. In fact, in the inset in Fig.~\ref{fig:PUEg}(a) we show on a linear relation between $F_U$ and $E_g$ on site I1, where the linear relation is independent of both \muI\ and \tildet.
%
To describe processes {\it along} the QSHI edge, it is therefore possible to use a model where the effect of the SC is only taken into account through an effective $\Delta_U = U_{\rm eff}F_U$ for the edge state, where the effective pairing potential $U_{\rm eff}$ is independent of both \muI\ and \tildet. Such an effective model has been completely dominating in the literature \cite{Fu08,Fu09,MajoranaTIsall},
and our calculations provide a fully self-consistent evidence for the validity of this approach. 
%
\section{Induced $p$-wave order parameters}
As established by Fu and Kane \cite{Fu08}, the low-energy spectrum in a 3D TI with an \swave\ superconducting order parameter resembles that of a spinless $p_x+ip_y$-wave SC in momentum space, although time-reversal symmetry is not broken. However, in the present case of a 2D QSHI with a 1D edge, any odd-in-momentum component necessarily only has a simple \pwave\ symmetry. Even though this \pwave\ symmetry is intimately linked to the generation of Majorana fermions, no detailed study has yet appeared on its origins. Stanescu {\it et al.} \cite{Stanescu10} very recently predicted, after some rather restrictive approximations, that the $p$-wave component is directly proportional to $\mu_{\rm I} \tilde{t}^2$, but no detailed calculations were preformed.
%
Here we explicitly calculate both on-site and bond superconducting order parameters that are odd in momentum, or \pwave. Since the $s$-wave state in the SC is induced through a local pairing and the QSHI edge state is extremely localized at the interface, all significant \pwave\ pairing amplitudes should be captured within these two, locally defined, quantities.
%

\begin{figure}[htb]
\includegraphics[scale = 0.85]{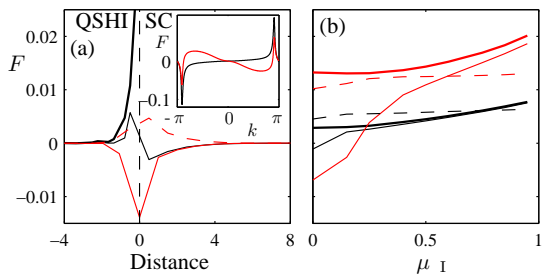}
\caption{\label{fig:Podd} (Color online) (a): $F_U$ (thick black), $F_{U}^o$ (thin black), $F_{J1}^o$ (thin red), and $F_{J2}^o = F_{J3}^o$ (thin dashed red) across the QSHI-SC interface for $\mu_{\rm I} = \tilde{t} = 0.5$. Vertical dashed line marks the interface and the distance is measured in number of unit cells. Inset: $k$-dependence at site I1 for $F_{Uk}^o$ (black) and $F_{J1k}^o$ (red).
(b): $F_U$ (thick) and $F_{U}^o$ (thin) on site I1 and $F_{J}^o$ (thin dashed) on the  \tildet-bond for $\tilde{t} = 0.2$ (black) and 0.6 (red) as a function of \muI.
}
\end{figure}

We start by defining the momentum-resolved on-site pairing amplitude $F_{Uk}(i) =  \langle a_{i-k\downarrow} a_{ik\uparrow}\rangle$. The conventional on-site order parameter can then be written as $F_U(i) = \frac{1}{N_k}\sum_{k = -\pi}^\pi F_{Uk}(i)$, where $N_k$ is the number of $k$-points. Obviously, only the part of $F_{Uk}$ that is even in $k$ contributes to $F_U$. There can also, technically, be a term in $F_{Uk}$ that is {\it odd} in $k$, which we will call $F_{Uk}^o$. For $\lambda = 0$ this odd component is zero, but on the edge of a QSHI it is finite.
We can understand this odd component better by defining the {\it spinless on-site} operator $c_{k} = \frac{1}{\sqrt{2}}(a_{k\uparrow} + a_{k\downarrow}$) at the QSHI edge. Then the Cooper pair amplitude $\langle c_{k}c_{-k}\rangle = \frac{1}{2} \langle a_{k\downarrow}a_{-k\uparrow} - a_{-k\downarrow}a_{k\uparrow}\rangle = F_{Uk}^o$, due to the explicit coupling of spin and momentum at the edge of a QSHI. Thus, $F_{U}^o = \frac{1}{N_k/2}\sum_{k = 0}^\pi F_{Uk}^o$ is nothing else than a spinless on-site $p$-wave pairing amplitude. In Fig.~\ref{fig:Podd}(a) we show the spatial extent of $F_{U}^o$ for a prototypical interface. We see that $F_{U}^o$ peaks at site I1 but that it is also present on site S1, and that there is a sign change between the two sites. 
This sign change interestingly also encodes an odd character {\it across} the interface, and we speculate that this might be a 1D manifestation of the $p_x + ip_y$-wave present on the surface of 3D TIs.
Fig.~\ref{fig:Podd}(b) furthermore shows that $F_{U}^o$ increases approximately linearly with \muI, with \tildet\ determining the slope. Since $F_{U}^o$ changes sign with $\mu_{\rm I}$, the finite charge transfer between the QSHI and the SC discussed above makes $F_{U}^o$ slightly negative for small positive (bulk) \muI\ at, especially, large \tildet. %
Without this charge transfer $F_{U}^o$ is zero at $\mu_{\rm I} = 0$. This can also be seen from the inset in Fig.~\ref{fig:Podd}(a), where we plot the $k$-dependence of the \pwave\ order parameters. $F_{Uk}^o$ peaks sharply at the $k$ value where we find the QSHI edge Dirac cone in the Bogoliubov spectrum. Since this cone is at $k = \pm \pi$ when $\mu_{\rm I} = 0$, time reversal symmetry will then force $F_{U}^o$ to vanish \cite{Fu08,Stanescu10}. We also note that for all small to moderately large tunneling amplitudes $F_{U}^o$ is linear in \tildet, which is also different from the earlier prediction of $\tilde{t}^2$ behavior \cite{Stanescu10}.
We also note that $F_{U}^o \leq F_{U}$ for all \muI\ and \tildet. This further supports our earlier conclusion that the induced \pwave\ state will never close the superconducting gap in the QSHI, even at high doping levels.

We also study the Cooper pair amplitude on nearest neighbor bonds, $F_{J\delta}(i) = \frac{1}{\sqrt{2}}\langle a_{i\downarrow}a_{i+\delta\uparrow} - a_{i\uparrow}a_{i+\delta\downarrow}\rangle$, where $\delta$ labels the bonds as defined in Fig.~\ref{fig:setup}. 
A small, but finite, $F_{J\delta}$ exists at the QSHI-SC interface, but its value is largely independent on $\lambda$. Thus this (even) order parameter is not a consequence of the topological nature of the QSHI but it is a property of the surface of the SC. 
%
However, by using the same momentum-resolved technique presented above for $F_U$, we can also define an odd-in-momentum component $F_{J\delta}^o$, which is only non-zero for finite $\lambda$. We find that especially $F_{J1}^o$(I1), which resides on the \tildet\ bond, is large. This could have been expected, since the QSHI edge state leaks substantially  into the SC.
Furthermore, we find that $F_{J\delta}^o$ increases approximately linearly with \tildet\ but that it is essentially {\it independent} on \muI, as seen in Fig.~\ref{fig:Podd}. Previous work \cite{Fu08,Stanescu10} has established a finite \pwave\ order parameter at high doping levels, but our results explicitly show that there also exists a finite induced \pwave\ order parameter at $\mu_{\rm I} = 0$. This is despite time-reversal symmetry dictating that $F_{J\delta k}^o = 0$ at $k = \pm \pi$ for half-filling. The finite $F_{J\delta}^o$ is still possible due to a rather broad $k$-dependence, as seen in the inset in Fig.~\ref{fig:Podd}(a).
%
To further analyze $F_{J\delta}^o$ we define the {\it spinless bond} operator $d_{k} = \frac{1}{2}[(a_{k\uparrow} - b_{k\uparrow}) - (a_{k\downarrow} - b_{k\downarrow})]$, where $a$ is on the I1 site and b is on the S1 site. Then we can construct the spinless Cooper pair amplitude residing on the \tildet\ bond as $\langle d_{k}d_{-k}\rangle = \frac{1}{2}[F_{U}^o({\rm I1}) + F_{U}^o({\rm S1}) - \sqrt{2}F_{J1}^o({\rm I1})] \equiv F_{J}^o$. Since $F_{U}^o$ at sites I1 and S1 have different signs, $F_{J1}^o$(I1) contributes crucially to this spinless \pwave\ bond order parameter.
In Fig.~\ref{fig:Podd}(b) wee see that $F_{J}^o$ increases moderately with \tildet, similar to $F_{U}$ and $F_{U}^o$, but, contrary to $F_{U}^o$, it is essentially independent on \muI\ in the whole range $0\leq \mu_{\rm I} < 1$. 
In summary, we have here explicitly shown that spinless \pwave\ order parameters exists for all doping levels of the QSHI. At low to moderate doping levels the spinless bond order parameter  $F_{J}^o$ dominates, whereas for very high doping levels the spinless on-site $F_{U}^o$ eventually becomes larger. This helps explain why the Majorana mode found in e.g.~vortex cores exists for all doping levels of the TI. 
%
%
%
\section{\dxywave\ superconductor}
While the \pwave\ character of an induced \swave\ superconducting state at a TI surface is of wide interest due to Majorana fermion creation, so are of course other systems with Majorana fermions. Recently Linder {\it et al.}~\cite{Linder10} discovered that Majorana fermions also exists at the surface of a \dxywave\ superconducting 3D TI.
The surface/interface of a \dxywave\ SC contains zero energy states, or midgap Andreev bound states, since $\Delta(\theta) = -\Delta(\pi-\theta)$, where $\theta$ is the angle of incidence (see e.g.~Ref.~\onlinecite{Kashiwaya00}). When these states exist on the surface of a TI, their double spin-degeneracy is broken and a Majorana fermion appears instead. This is contrast to the topologically trivial high-T$_c$ SCs where spin-degeneracy is never broken but gives rise to "double" Majorana fermions.
From an experimental perspective, however, there are two crucially important questions that needs to be addressed before realizing this Majorana system.
(1): In a self-consistent model of a QSHI - $d_{xy}$ SC interface, will the induced superconducting state in the QSHI be of the required \dxywave\ character, and (2): how protected is the Majorana fermion, i.e.~how large is the excitation gap? 

\begin{figure}[htb]
\includegraphics[scale = 0.85]{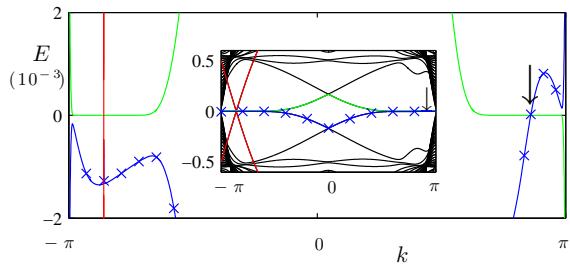}
\caption{\label{fig:dwave} (Color online) Bogoliubov band structure for the QSHI - $d_{xy}$ SC structure with $\mu_{\rm I} = \tilde{t} =0.5$ and $J = 1$. Unaffected by the interface are the Dirac cone on left hand side of the QSHI (red, close to $k=-\pi$) and Andreev bound states at the right hand side of the SC (green). Andreev bound states at the QSHI-SC interface (blue, crosses) (resolution in $k$ much larger than crosses indicate) with a single Majorana fermion mode (arrow). Inset: zoom-out at higher energies. Conical structure at $k=\pm \pi$ is due to the nodes in the \dxywave\ order parameter.
}
\end{figure}

By rotating the SC in Fig.~\ref{fig:setup} to a (1,1) interface and, instead of $U$, using an effective pairing amplitude $J $ on nearest neighbor bonds, we can model an interface between a QSHI and a \dxywave\ SC. We start the self-consistency loop with $\Delta_{J\delta} = -J F_{J\delta}$ having a pure \dxywave\ symmetry, i.e. $\Delta_{J1} = - \Delta_{J2}$, but we do not impose any symmetry constraints on $\Delta_{J\delta}$ during the loop, as to not prevent the formation of other, competing pairing symmetries at the interface. 
This procedure results in a solution that is fully \dxywave\ in the bulk and persists out to the S1 site with only a moderately large inverse proximity effect. This \dxywave\ state also leaks weakly into the QSHI insulator where it very rapidly decays. However, there is a larger, but imaginary, component of $F_{J1}$ on the \tildet-bond, accompanied by an imaginary and opposite signed $F_{J2} = F_{J3}$. In aggregate this gives a local $d_{xy}+id_{x^2-y^2}$-wave symmetry \cite{graphenedwave} at the QSHI edge. 
Thus an effective model with a purely \dxywave\ superconducting order parameter at the QSHI edge is not valid for a proximity induced \dxywave\ state. However, a small, but finite, \dxywave\ component still exists at the interface, so zero energy states might still exist at the interface. We thus turn to investigating the Bogoliubov band structure in Fig.~\ref{fig:dwave}. 
%
The Dirac cone (red) close to $k=-\pi$ (exact position is set by \muI) lives on the left hand side of the QSHI and is thus unaffected by the SC. The low-lying band at $E\geq 0$ (green) lives on the unaffacted right hand side of the SC and thus gives the doubly degenerate zero energy states always present at the surface of \dxywave\ SCs. 
For $\tilde{t} = 0$, the zero energy states on the left hand side of the SC creates a mirror-symmetric band at $E\leq 0$ to the green band. However, for finite \tildet\ this band (blue, crosses) is found mainly at non-zero energies and it has only {\it one single} crossing of $E = 0$ (arrow). This is a single Majorana fermion mode and its $k$-value is the same as the $k$-value for the (now gapped) Dirac apex of the QSHI edge state. 
However, this mode joins the nodal quasiparticle excitations arising from the bulk $d$-wave order parameter at $k = \pm \pi$. Thus, this single Majorana mode has {\it no} excitation gap to topologically trivial states.
Thus, the single Majorana fermion at $E = 0$ has {\it no} excitation gap to topologically trivial states. 
The band structure in Fig.~\ref{fig:dwave} is qualitatively similar for all finite \tildet\ and moderately small to large \muI. However, for $\mu_I \approx 0$ the would-be Majorana mode merges with the conical \dwave\ nodal structure at $k=\pm \pi$ and disappears, i.e.~there is then no state at zero energy that is localized at the QSHI-SC interface. 
We also note that in order to achieve the band structure in Fig.~\ref{fig:dwave} self-consistency is crucial. With a step-function order parameter $\Delta_J$, the band for the zero energy states at the QSHI-SC interface is essentially mirror symmetric to the equivalent band residing on the unaffected side of the SC, and thus no single Majorana mode is created.
This is in sharp contrast to our results for a \swave\ SC and is due to the intricate interplay between the QSHI edge state and the zero energy states of the \dxywave\ SC.
To summarize this section, we have shown that a single Majorana fermion exists only for doped QSHI-\dxywave\ SC structures and that there is no excitation gap for this mode. Thus, from an experimental point of view, we do not expect this Majorana fermion to be stable for quantum computing operations \cite{Nayak08, Sau10}.

%
\begin{acknowledgments}
The author thanks Jacob Linder and Eddy Ardonne for very valuable discussions.
\end{acknowledgments}


\begin{thebibliography}{24}
\expandafter\ifx\csname natexlab\endcsname\relax\def\natexlab#1{#1}\fi
\expandafter\ifx\csname bibnamefont\endcsname\relax
  \def\bibnamefont#1{#1}\fi
\expandafter\ifx\csname bibfnamefont\endcsname\relax
  \def\bibfnamefont#1{#1}\fi
\expandafter\ifx\csname citenamefont\endcsname\relax
  \def\citenamefont#1{#1}\fi
\expandafter\ifx\csname url\endcsname\relax
  \def\url#1{\texttt{#1}}\fi
\expandafter\ifx\csname urlprefix\endcsname\relax\def\urlprefix{URL }\fi
\providecommand{\bibinfo}[2]{#2}
\providecommand{\eprint}[2][]{\url{#2}}

\bibitem[{\citenamefont{Bernevig et~al.}(2006)\citenamefont{Bernevig, Hughes,
  and Zhang}}]{Bernevig06all}
\bibinfo{author}{\bibfnamefont{B.~A.} \bibnamefont{Bernevig}},
  \bibinfo{author}{\bibfnamefont{T.~L.} \bibnamefont{Hughes}},
  \bibnamefont{and} \bibinfo{author}{\bibfnamefont{S.-C.} \bibnamefont{Zhang}},
  \bibinfo{journal}{Science} \textbf{\bibinfo{volume}{314}},
  \bibinfo{pages}{1757} (\bibinfo{year}{2006});
\bibinfo{author}{\bibfnamefont{B.~A.} \bibnamefont{Bernevig}} \bibnamefont{and}
  \bibinfo{author}{\bibfnamefont{S.-C.} \bibnamefont{Zhang}},
  \bibinfo{journal}{Phys. Rev. Lett.} \textbf{\bibinfo{volume}{96}},
  \bibinfo{pages}{106802} (\bibinfo{year}{2006}).

\bibitem[{\citenamefont{Konig et~al.}(2007)\citenamefont{Konig, Wiedmann,
  Brune, Roth, Buhmann, Molenkamp, Qi, and Zhang}}]{Konig07}
\bibinfo{author}{\bibfnamefont{M.}~\bibnamefont{Konig}},
  \bibinfo{author}{\bibfnamefont{S.}~\bibnamefont{Wiedmann}},
  \bibinfo{author}{\bibfnamefont{C.}~\bibnamefont{Brune}},
  \bibinfo{author}{\bibfnamefont{A.}~\bibnamefont{Roth}},
  \bibinfo{author}{\bibfnamefont{H.}~\bibnamefont{Buhmann}},
  \bibinfo{author}{\bibfnamefont{L.~W.} \bibnamefont{Molenkamp}},
  \bibinfo{author}{\bibfnamefont{X.-L.} \bibnamefont{Qi}}, \bibnamefont{and}
  \bibinfo{author}{\bibfnamefont{S.-C.} \bibnamefont{Zhang}},
  \bibinfo{journal}{Science} \textbf{\bibinfo{volume}{318}},
  \bibinfo{pages}{766} (\bibinfo{year}{2007}).

\bibitem[{\citenamefont{Hsieh et~al.}(2008)\citenamefont{Hsieh, Qian, Wray,
  Xia, Hor, Cava, and Hasan}}]{Hsieh08all}
\bibinfo{author}{\bibfnamefont{D.}~\bibnamefont{Hsieh}},
  \bibinfo{author}{\bibfnamefont{D.}~\bibnamefont{Qian}},
  \bibinfo{author}{\bibfnamefont{L.}~\bibnamefont{Wray}},
  \bibinfo{author}{\bibfnamefont{Y.}~\bibnamefont{Xia}},
  \bibinfo{author}{\bibfnamefont{Y.~S.} \bibnamefont{Hor}},
  \bibinfo{author}{\bibfnamefont{R.~J.} \bibnamefont{Cava}}, \bibnamefont{and}
  \bibinfo{author}{\bibfnamefont{M.~Z.} \bibnamefont{Hasan}},
  \bibinfo{journal}{Nature} \textbf{\bibinfo{volume}{452}},
  \bibinfo{pages}{970} (\bibinfo{year}{2008});
\bibinfo{author}{\bibfnamefont{D.}~\bibnamefont{Hsieh}},
  \bibinfo{author}{\bibfnamefont{Y.}~\bibnamefont{Xia}},
  \bibinfo{author}{\bibfnamefont{L.}~\bibnamefont{Wray}},
  \bibinfo{author}{\bibfnamefont{D.}~\bibnamefont{Qian}},
  \bibinfo{author}{\bibfnamefont{A.}~\bibnamefont{Pal}},
  \bibinfo{author}{\bibfnamefont{J.~H.} \bibnamefont{Dil}},
  \bibinfo{author}{\bibfnamefont{J.}~\bibnamefont{Osterwalder}},
  \bibinfo{author}{\bibfnamefont{F.}~\bibnamefont{Meier}},
  \bibinfo{author}{\bibfnamefont{G.}~\bibnamefont{Bihlmayer}},
  \bibinfo{author}{\bibfnamefont{C.~L.} \bibnamefont{Kane}},
  \bibnamefont{et~al.}, \bibinfo{journal}{Science}
  \textbf{\bibinfo{volume}{323}}, \bibinfo{pages}{919} (\bibinfo{year}{2009}).

\bibitem[{\citenamefont{Hasan and Kane}()}]{Hasan10}
\bibinfo{author}{\bibfnamefont{M.~Z.} \bibnamefont{Hasan}} \bibnamefont{and}
  \bibinfo{author}{\bibfnamefont{C.~L.} \bibnamefont{Kane}},
 \bibinfo{journal}{Rev. Mod. Phys.} \textbf{\bibinfo{volume}{82}},
  \bibinfo{pages}{3045} (\bibinfo{year}{2010}).

\bibitem[{\citenamefont{Qi and Zhang}()}]{Qi10}
\bibinfo{author}{\bibfnamefont{X.-L.} \bibnamefont{Qi}} \bibnamefont{and}
  \bibinfo{author}{\bibfnamefont{S.-C.} \bibnamefont{Zhang}},
  \eprint{arXiv:1008.2026 (unpublished)}.

\bibitem[{\citenamefont{Fu and Kane}(2008)}]{Fu08}
\bibinfo{author}{\bibfnamefont{L.}~\bibnamefont{Fu}} \bibnamefont{and}
  \bibinfo{author}{\bibfnamefont{C.~L.} \bibnamefont{Kane}},
  \bibinfo{journal}{Phys. Rev. Lett.} \textbf{\bibinfo{volume}{100}},
  \bibinfo{pages}{096407} (\bibinfo{year}{2008}).

\bibitem[{\citenamefont{Read and Green}(2000)}]{Read00}
\bibinfo{author}{\bibfnamefont{N.}~\bibnamefont{Read}} \bibnamefont{and}
  \bibinfo{author}{\bibfnamefont{D.}~\bibnamefont{Green}},
  \bibinfo{journal}{Phys. Rev. B} \textbf{\bibinfo{volume}{61}},
  \bibinfo{pages}{10267} (\bibinfo{year}{2000}).

\bibitem[{\citenamefont{Fu and Kane}(2009{\natexlab{a}})}]{Fu09}
\bibinfo{author}{\bibfnamefont{L.}~\bibnamefont{Fu}} \bibnamefont{and}
  \bibinfo{author}{\bibfnamefont{C.~L.} \bibnamefont{Kane}},
  \bibinfo{journal}{Phys. Rev. B} \textbf{\bibinfo{volume}{79}},
  \bibinfo{pages}{161408} (\bibinfo{year}{2009}{\natexlab{a}}).

\bibitem[{\citenamefont{Nayak et~al.}(2008)\citenamefont{Nayak, Simon, Stern,
  Freedman, and Das~Sarma}}]{Nayak08}
\bibinfo{author}{\bibfnamefont{C.}~\bibnamefont{Nayak}},
  \bibinfo{author}{\bibfnamefont{S.~H.} \bibnamefont{Simon}},
  \bibinfo{author}{\bibfnamefont{A.}~\bibnamefont{Stern}},
  \bibinfo{author}{\bibfnamefont{M.}~\bibnamefont{Freedman}}, \bibnamefont{and}
  \bibinfo{author}{\bibfnamefont{S.}~\bibnamefont{Das~Sarma}},
  \bibinfo{journal}{Rev. Mod. Phys.} \textbf{\bibinfo{volume}{80}},
  \bibinfo{pages}{1083} (\bibinfo{year}{2008}).

\bibitem[{\citenamefont{Nilsson et~al.}(2008)\citenamefont{Nilsson, Akhmerov,
  and Beenakker}}]{MajoranaTIsall}
\bibinfo{author}{\bibfnamefont{J.}~\bibnamefont{Nilsson}},
  \bibinfo{author}{\bibfnamefont{A.~R.} \bibnamefont{Akhmerov}},
  \bibnamefont{and} \bibinfo{author}{\bibfnamefont{C.~W.~J.}
  \bibnamefont{Beenakker}}, \bibinfo{journal}{Phys. Rev. Lett.}
  \textbf{\bibinfo{volume}{101}}, \bibinfo{pages}{120403}
  (\bibinfo{year}{2008});
\bibinfo{author}{\bibfnamefont{A.~R.} \bibnamefont{Akhmerov}},
  \bibinfo{author}{\bibfnamefont{J.}~\bibnamefont{Nilsson}}, \bibnamefont{and}
  \bibinfo{author}{\bibfnamefont{C.~W.~J.} \bibnamefont{Beenakker}},
  \bibinfo{journal}{Phys. Rev. Lett.} \textbf{\bibinfo{volume}{102}},
  \bibinfo{pages}{216404} (\bibinfo{year}{2009});
\bibinfo{author}{\bibfnamefont{P.}~\bibnamefont{Adroguer}},
  \bibinfo{author}{\bibfnamefont{C.}~\bibnamefont{Grenier}},
  \bibinfo{author}{\bibfnamefont{D.}~\bibnamefont{Carpentier}},
  \bibinfo{author}{\bibfnamefont{J.}~\bibnamefont{Cayssol}},
  \bibinfo{author}{\bibfnamefont{P.}~\bibnamefont{Degiovanni}},
  \bibnamefont{and} \bibinfo{author}{\bibfnamefont{E.}~\bibnamefont{Orignac}},
  \bibinfo{journal}{Phys. Rev. B} \textbf{\bibinfo{volume}{82}},
  \bibinfo{pages}{081303} (\bibinfo{year}{2010});
\bibinfo{author}{\bibfnamefont{Y.}~\bibnamefont{Tanaka}},
  \bibinfo{author}{\bibfnamefont{T.}~\bibnamefont{Yokoyama}}, \bibnamefont{and}
  \bibinfo{author}{\bibfnamefont{N.}~\bibnamefont{Nagaosa}},
  \bibinfo{journal}{Phys. Rev. Lett.} \textbf{\bibinfo{volume}{103}},
  \bibinfo{pages}{107002} (\bibinfo{year}{2009});
\bibinfo{author}{\bibfnamefont{L.}~\bibnamefont{Fu}} \bibnamefont{and}
  \bibinfo{author}{\bibfnamefont{C.~L.} \bibnamefont{Kane}},
  \bibinfo{journal}{Phys. Rev. Lett.} \textbf{\bibinfo{volume}{102}},
  \bibinfo{pages}{216403} (\bibinfo{year}{2009}{\natexlab{b}}).

\bibitem[{\citenamefont{Stanescu et~al.}(2010)\citenamefont{Stanescu, Sau,
  Lutchyn, and Das~Sarma}}]{Stanescu10}
\bibinfo{author}{\bibfnamefont{T.~D.} \bibnamefont{Stanescu}},
  \bibinfo{author}{\bibfnamefont{J.~D.} \bibnamefont{Sau}},
  \bibinfo{author}{\bibfnamefont{R.~M.} \bibnamefont{Lutchyn}},
  \bibnamefont{and}
  \bibinfo{author}{\bibfnamefont{S.}~\bibnamefont{Das~Sarma}},
  \bibinfo{journal}{Phys. Rev. B} \textbf{\bibinfo{volume}{81}},
  \bibinfo{pages}{241310(R)} (\bibinfo{year}{2010}).

\bibitem[{\citenamefont{Linder et~al.}(2010)\citenamefont{Linder, Tanaka,
  Yokoyama, Sudb\o{}, and Nagaosa}}]{Linder10}
\bibinfo{author}{\bibfnamefont{J.}~\bibnamefont{Linder}},
  \bibinfo{author}{\bibfnamefont{Y.}~\bibnamefont{Tanaka}},
  \bibinfo{author}{\bibfnamefont{T.}~\bibnamefont{Yokoyama}},
  \bibinfo{author}{\bibfnamefont{A.}~\bibnamefont{Sudb\o{}}}, \bibnamefont{and}
  \bibinfo{author}{\bibfnamefont{N.}~\bibnamefont{Nagaosa}},
  \bibinfo{journal}{Phys. Rev. Lett.} \textbf{\bibinfo{volume}{104}},
  \bibinfo{pages}{067001} (\bibinfo{year}{2010}).

\bibitem[{\citenamefont{Sau et~al.}(2010)\citenamefont{Sau, Lutchyn, Tewari,
  and Das~Sarma}}]{Sau10}
\bibinfo{author}{\bibfnamefont{J.~D.} \bibnamefont{Sau}},
  \bibinfo{author}{\bibfnamefont{R.~M.} \bibnamefont{Lutchyn}},
  \bibinfo{author}{\bibfnamefont{S.}~\bibnamefont{Tewari}}, \bibnamefont{and}
  \bibinfo{author}{\bibfnamefont{S.}~\bibnamefont{Das~Sarma}},
  \bibinfo{journal}{Phys. Rev. B} \textbf{\bibinfo{volume}{82}},
  \bibinfo{pages}{094522} (\bibinfo{year}{2010}).

\bibitem[{\citenamefont{Kane and Mele}(2005{\natexlab{a}})}]{KaneMele05}
\bibinfo{author}{\bibfnamefont{C.~L.} \bibnamefont{Kane}} \bibnamefont{and}
  \bibinfo{author}{\bibfnamefont{E.~J.} \bibnamefont{Mele}},
  \bibinfo{journal}{Phys. Rev. Lett.} \textbf{\bibinfo{volume}{95}},
  \bibinfo{pages}{226801} (\bibinfo{year}{2005}{\natexlab{a}});
\bibinfo{author}{\bibfnamefont{C.~L.} \bibnamefont{Kane}} \bibnamefont{and}
  \bibinfo{author}{\bibfnamefont{E.~J.} \bibnamefont{Mele}},
  \bibinfo{journal}{Phys. Rev. Lett.} \textbf{\bibinfo{volume}{95}},
  \bibinfo{pages}{146802} (\bibinfo{year}{2005}{\natexlab{b}}).

\bibitem[{\citenamefont{Kashiwaya and Tanaka}(2000)}]{Kashiwaya00}
\bibinfo{author}{\bibfnamefont{S.}~\bibnamefont{Kashiwaya}} \bibnamefont{and}
  \bibinfo{author}{\bibfnamefont{Y.}~\bibnamefont{Tanaka}},
  \bibinfo{journal}{Rep.\ Prog.\ Phys.} \textbf{\bibinfo{volume}{63}},
  \bibinfo{pages}{1641} (\bibinfo{year}{2000}).

\bibitem[{\citenamefont{Black-Schaffer and Doniach}(2007)}]{graphenedwave}
\bibinfo{author}{\bibfnamefont{A.~M.} \bibnamefont{Black-Schaffer}}
  \bibnamefont{and} \bibinfo{author}{\bibfnamefont{S.}~\bibnamefont{Doniach}},
  \bibinfo{journal}{Phys.\ Rev.\ B} \textbf{\bibinfo{volume}{75}},
  \bibinfo{pages}{134512} (\bibinfo{year}{2007});
\bibinfo{author}{\bibfnamefont{J.}~\bibnamefont{Linder}},
  \bibinfo{author}{\bibfnamefont{A.~M.} \bibnamefont{Black-Schaffer}},
  \bibinfo{author}{\bibfnamefont{T.}~\bibnamefont{Yokoyama}},
  \bibinfo{author}{\bibfnamefont{S.}~\bibnamefont{Doniach}}, \bibnamefont{and}
  \bibinfo{author}{\bibfnamefont{A.}~\bibnamefont{Sudb\o{}}},
  \bibinfo{journal}{Phys. Rev. B} \textbf{\bibinfo{volume}{80}},
  \bibinfo{pages}{094522} (\bibinfo{year}{2009}).

\end{thebibliography}

\end{document}